\documentclass[10pt]{article}

\usepackage[margin=1in,footskip=0.25in]{geometry}
\usepackage{lineno,hyperref}
\usepackage{amsmath}
\usepackage{amsfonts,mathtools}
\modulolinenumbers[5]
\usepackage{array}
\usepackage{subfig}
\usepackage{gensymb}
\usepackage{xcolor}
\usepackage{empheq}
\usepackage[many]{tcolorbox}
\usepackage{titlesec}
\titlelabel{\thetitle.\ }
\usepackage[symbol]{footmisc}
\usepackage[linesnumbered,ruled,vlined]{algorithm2e}
\usepackage{morefloats}
\usepackage{mathrsfs}
\usepackage{relsize}

\makeatletter
\newcommand{\mathleft}{\@fleqntrue\@mathmargin\parindent}
\newcommand{\mathcenter}{\@fleqnfalse}
\makeatother
\usepackage{cite}

\newcommand*{\affaddr}[1]{#1}
\newcommand*{\affmark}[1][*]{\textsuperscript{#1}}

\def\correspondingauthor{\footnote{Corresponding author: Reza Behrou (rbehrou@jhu.edu)}}

\providecommand{\keywords}[1]{\textbf{\textbf{\small Keywords:}} #1}

\begin{document}

%
\date{}

\title{\normalsize \bfseries Topology optimization for transient response of structures subjected to dynamic loads}

\author{%
\small Reza Behrou\affmark[1,]\correspondingauthor{} \ and James K. Guest\affmark[1] \\
\affaddr{\affmark[1]\footnotesize Department of Civil Engineering, Johns Hopkins University, Baltimore, MD, 21218, USA}%
}

\maketitle

%
\begin{abstract}
\label{sec:Abstract}
This paper presents a topology optimization framework for structural problems subjected to transient loading. The mechanical model assumes a linear elastic isotropic material, infinitesimal strains, and a dynamic response. The optimization problem is solved using the gradient-based optimizer Method of Moving Asymptotes (MMA) with time-dependent sensitivities provided via the adjoint method. The stiffness of materials is interpolated using the Solid Isotropic Material with Penalization (SIMP) method and the Heaviside Projection Method (HPM) is used to stabilize the problem numerically and improve the manufacturability of the topology-optimized designs. Both static and dynamic optimization examples are considered here. The resulting optimized designs demonstrate the ability of topology optimization to tailor the transient response of structures. 
\end{abstract}


\section*{Nomenclature}

\begin{tabbing}
  XXX \= \kill
  $\boldsymbol{M}$                      \> Mass matrix \\
  $\boldsymbol{C}$                      \> Damping matrix \\
  $\boldsymbol{K}$                      \> Stiffness matrix \\
  $\boldsymbol{\ddot u}$                \> Acceleration vector \\
  $\boldsymbol{\dot u}$                 \> Velocity vector \\
  $\boldsymbol{u}$                      \> Displacement vector \\
  $\boldsymbol{u}_{\mathrm{int}}$       \> Initial solution vector \\
  $\boldsymbol{\bar f}$                 \> External force vector \\
  $\boldsymbol{R}$                      \> Residual vector \\
  $\boldsymbol{\phi}$                   \> Vector of design variables \\
  $\boldsymbol{\rho}$                   \> Vector of densities \\
  $\boldsymbol{\lambda}$                \> Vector of adjoint variables \\
  $\boldsymbol{I}$                      \> Identity matrix \\
  $Z$                                   \> Total objective function \\
  $H_{i}$                               \> Equality constraint \\
  $G_{j}$                               \> Inequality constraint \\
  $\phi_{min}$                          \> Lower bound of the design variables \\
  $\phi_{max}$                          \> Upper bound of the design variables \\
  $N_{t}$                               \> Number of time steps \\
  $N_{h}$                               \> Number of equality constraints \\
  $N_{g}$                               \> Number of inequality constraints \\
  $N_{\phi}$                            \> Number of design variables \\
  $\rho^{e}$                            \> Elemental density \\
  $\eta$                                \> SIMP exponent penalty \\
  $\rho^{e}_{min}$                      \> Small positive number \\
  $E_{0}$                               \> Young's modulus of pure solid \\ 
  $r_{min}$                             \> Minimum desired feature size \\
  $\beta$                               \> Curvature of regularization \\
  $V_{max}$                             \> Maximum allowable volume fraction \\
  $\tilde \beta$                        \> Newmark parameter \\
  $\tilde \gamma$                       \> Newmark parameter \\
  $\Delta t$                            \> Time step size  \\[5pt]
  \textit{Superscript}\\
  $n$                                   \> Current time \\
 \end{tabbing}
%

%
\section{Introduction
        \label{sec:Introduction}}
Topology optimization as a form-finding methodology for design has developed rapidly during recent years. Its applications cover different fields of engineering such as structural mechanics, fluid flow, heat transfer problems, and so on \cite{BS:03, BP:03, OG:16, BLM:17, LZG+:15, AAA+:14}. For structural problems, it has a demonstrated capability of identifying novel and high performance structures when the operating environment is consistent with assumed load cases used in the optimization. The vast majority of recent studies, however, assume these load cases are static and deterministic. Although significant progress has recently been made considering stochastic dynamic loads described by power spectra, see for example \cite{ZYG+:17}, the area of topology optimization under  dynamic loads, including the challenging problem of transient loading, is less studied. Some progress has been made in optimization for free vibration problems \cite{DO:07,KIKUCHI:92, SA:93} and forced vibrations \cite{MKH:93, MKC:95, Jensen:07}, including deterministic dynamic and transient loading through time history analysis \cite{KPA:06}. This paper presents a gradient-based topology optimization framework for structures subjected to time-dependent and transient loading. 

Optimizing the dynamic response of structures has many important applications in different fields of engineering, including structural, mechanical, and aerospace engineering. Analysis is often performed in the time or frequency domains, and topology optimization likewise follows these domains \cite{KPA:06,ZWR+:16}. The objectives of topology optimization in frequency domain are typically minimizing the frequency response of the structure with an excitation frequency or a frequency range (see for example \cite{SWF:11}), or minimizing the structural vibration under harmonic excitation (see for example \cite{ZK:13, LZG:15}). In the time domain, the objective is often defined as minimizing the strain energy in the structure (maximizing stiffness) within a specific time interval. Due to transient nature of applied loads, designing a structure that preserves the same functionality within different time ranges is a crucial factor in structural dynamics. In the time domain, topology optimization under dynamic loading is fairy difficult and the sensitivities can be costly because many time-dependent analyses are required. Most recent studies, therefore, have focused on simplified methods such as performing topology optimization under a system of equivalent static loads (see for example \cite{CP:02, KP:10, LP:15, JLL+:12}). The results of topology optimization are used in dynamic analysis of the structure to satisfy the convergence criteria \cite{JLL+:12}. Model reduction methods using mode displacement  and mode acceleration methods have been considered to simplify topology optimization under dynamic loading and reduce computation costs (see for example \cite{ZW:16}). Due to the simplifying assumptions of these methods, the engineer may not gain complete control or be able to optimize a specific property of the actual transient response under dynamic loads. For more information about dynamic optimization in time and frequency domains the reader is referred to \cite{Venini:16}. 

Gradient-free optimization methods have been applied  to many structural problems with dynamic and transient responses \cite{KEMB:02, SH:04, FLP:06}. Unlike  gradient-based optimization, these methods do not perform direct sensitivity analysis during the optimization process and instead either generate design alternatives randomly or probabilistically, or approximate design sensitivities using sampling points. These methods are most successful when dimensionality is relatively small, and generally not practical for large scale finite element models with large number of design variables. Gradient-based optimization, on the other hand, offers an effective approach to solve a broad range of topology optimization problems \cite{DG:14}. The computational complexities of sensitivity analysis, however, dramatically increase for transient problems (see for example \cite{MTV:94}). Additionally,  gradient-based optimization requires an interpolation and penalization scheme to create a continuous representation of material stiffness. The Solid Isotropic Material with Penalization (SIMP) method is the most popular of these methods and uses a power-law interpolation.  For more information the reader is referred to \cite{Rozvany:09}.

In this paper, we present a topology optimization framework for dynamic response of structures in time domain. The resulting transient topology optimization problem is solved using the gradient-based optimizer Method of Moving Asymptotes (MMA) \cite{Svanberg:87}, with time-dependent sensitivities provided via the adjoint method.  The SIMP method is used for interpolation of the stiffness of materials. The Heaviside Projection Method (HPM) \cite{GPB:04} is used to satisfy the desired minimum length scale of the topological features without adding additional constraints to the problem, thereby influencing manufacturability as well as circumventing well-known instabilities of checkerboard patterns and solution mesh dependency. 
%

%
\section{Physical model
        \label{sec:PhysicalModel}}
For elastodynamics problems, the time discretized form of the equation of motion is expressed as follows:
\begin{equation}\label{eq:EquationofMotion_1}
\begin{split}
\boldsymbol{M} \boldsymbol{\ddot u}^{n} + \boldsymbol{C} \boldsymbol{\dot u}^{n} + \boldsymbol{K} \boldsymbol{u}^{n} - \boldsymbol{\bar f}^{n} = \boldsymbol{0},
\end{split}
\end{equation}
where the superscript $"n"$ denotes current time, $\boldsymbol{M}$, $\boldsymbol{C}$ and $\boldsymbol{K}$ are the mass, viscous damping and stiffness matrices, respectively. In this paper, the viscous damping matrix, $\boldsymbol{C}$, is assumed to be zero. $\boldsymbol{\bar f}^{n}$ is the external force vector at current time, $\boldsymbol{\ddot u}$, $\boldsymbol{\dot u}$ and $\boldsymbol{u}$ are the acceleration, velocity and displacement vectors, respectively. Using the Newmark approach \cite{Newmark:59}, the accelerations and velocities are expressed as follows:
\begin{equation}\label{eq:EquationofMotion_2}
\begin{split}
\boldsymbol{\dot u}^{n} = \left( 1 - \frac{\tilde \gamma}{\tilde \beta} \right) \boldsymbol{\dot u}^{n-1} + \left( 1 - \frac{\tilde \gamma}{2\tilde \beta} \right) \Delta t \boldsymbol{\ddot u}^{n-1} + \frac{\tilde \gamma}{\tilde \beta \Delta t} \left( \boldsymbol{u}^{n} - \boldsymbol{u}^{n-1} \right),
\end{split}
\end{equation}
\begin{equation}\label{eq:EquationofMotion_3}
\begin{split}
\boldsymbol{\ddot u}^{n} = -\frac{1}{\tilde \beta \Delta t} \boldsymbol{\dot u}^{n-1} + \left( \frac{1}{2\tilde \beta} -1 \right) \boldsymbol{\ddot u}^{n-1} + \frac{1}{\tilde \beta \Delta t^{2}} \left( \boldsymbol{u}^{n} - \boldsymbol{u}^{n-1} \right),
\end{split}
\end{equation}
where $\tilde \beta = 0.25$ and $\tilde \gamma = 0.5$ are parameters for a particular member of the Newmark family \cite{JE:94}, and $\Delta t$ is the time step size. Substituting Equations \eqref{eq:EquationofMotion_2} and \eqref{eq:EquationofMotion_3} into \eqref{eq:EquationofMotion_1} yields the following residual equation for the elastodynamics problems:
\begin{equation}\label{eq:EquationofMotion_4}
\begin{split}
\boldsymbol{R}^{n} = \left [ \left( \frac{1}{2 \tilde \beta} -1 \right) \boldsymbol{M} + \left( 1 - \frac{\tilde \gamma}{2\tilde \beta} \right) \Delta t \boldsymbol{C} \right ] \boldsymbol{\ddot u}^{n-1} - \left [ \frac{1}{\tilde \beta \Delta t} \boldsymbol{M} - \left( 1 - \frac{\tilde \gamma}{\tilde \beta} \right) \boldsymbol{C} \right ] \boldsymbol{\dot u}^{n-1} \\
+ \left [ \frac{1}{\tilde \beta \Delta t^{2}} \boldsymbol{M} + \frac{\tilde \gamma}{\tilde \beta \Delta t} \boldsymbol{C} \right ] \left( \boldsymbol{u}^{n} - \boldsymbol{u}^{n-1} \right) + \boldsymbol{K} \boldsymbol{u}^{n} - \boldsymbol{\bar f}^{n} \quad \forall \quad n = 1, ..., N_{t}.
\end{split}
\end{equation}
This equation is solved for the displacement vector, $\boldsymbol{u}^{n}$, at time $t_{n}$ based on the displacement, $\boldsymbol{u}^{n-1}$, acceleration, $\boldsymbol{\ddot u}^{n-1}$, and velocity, $\boldsymbol{\dot u}^{n-1}$, vectors at the previous time, $t_{n-1}$. The acceleration, $\boldsymbol{\ddot u}^{n}$, and velocity, $\boldsymbol{\dot u}^{n}$, are updated recursively using Equations \eqref{eq:EquationofMotion_2} and \eqref{eq:EquationofMotion_3}. At the zeroth step, $n = 0$, we assume:
\begin{equation}\label{eq:EquationofMotion_4_1}
\begin{split}
\boldsymbol{R}^{0} = \boldsymbol{u}^{0} - \boldsymbol{u}_{\mathrm{int}},
\end{split}
\end{equation}
where $\boldsymbol{u}_{\mathrm{int}}$ is the initial condition assumed to be zero.
 
%
\section{Optimization problem
        \label{sec:OptimizationProblem}}
We present a topology optimization framework for structural problems under transient loading. A representative configuration of an optimization problem considered in this paper is shown in Figure \ref{fig:OptimizationProblem}. The design domain, $\Omega$, is subjected to a time-dependent loading condition applied at the boundary $\Gamma_{t}$ while displacements are prescribed at the boundary $\Gamma_{u}$. In the optimization examples presented in this paper, we seek to maximize the stiffness (minimize the strain energy) of the structure subjected to transient loading. For transient and dynamic responses, we define the generalized form of the optimization problem in a time-dependent discrete fashion as follows:
\begin{figure}[!b]
\centering
\subfloat[schematic of the optimization problem]{
  \includegraphics[width=0.40\linewidth]{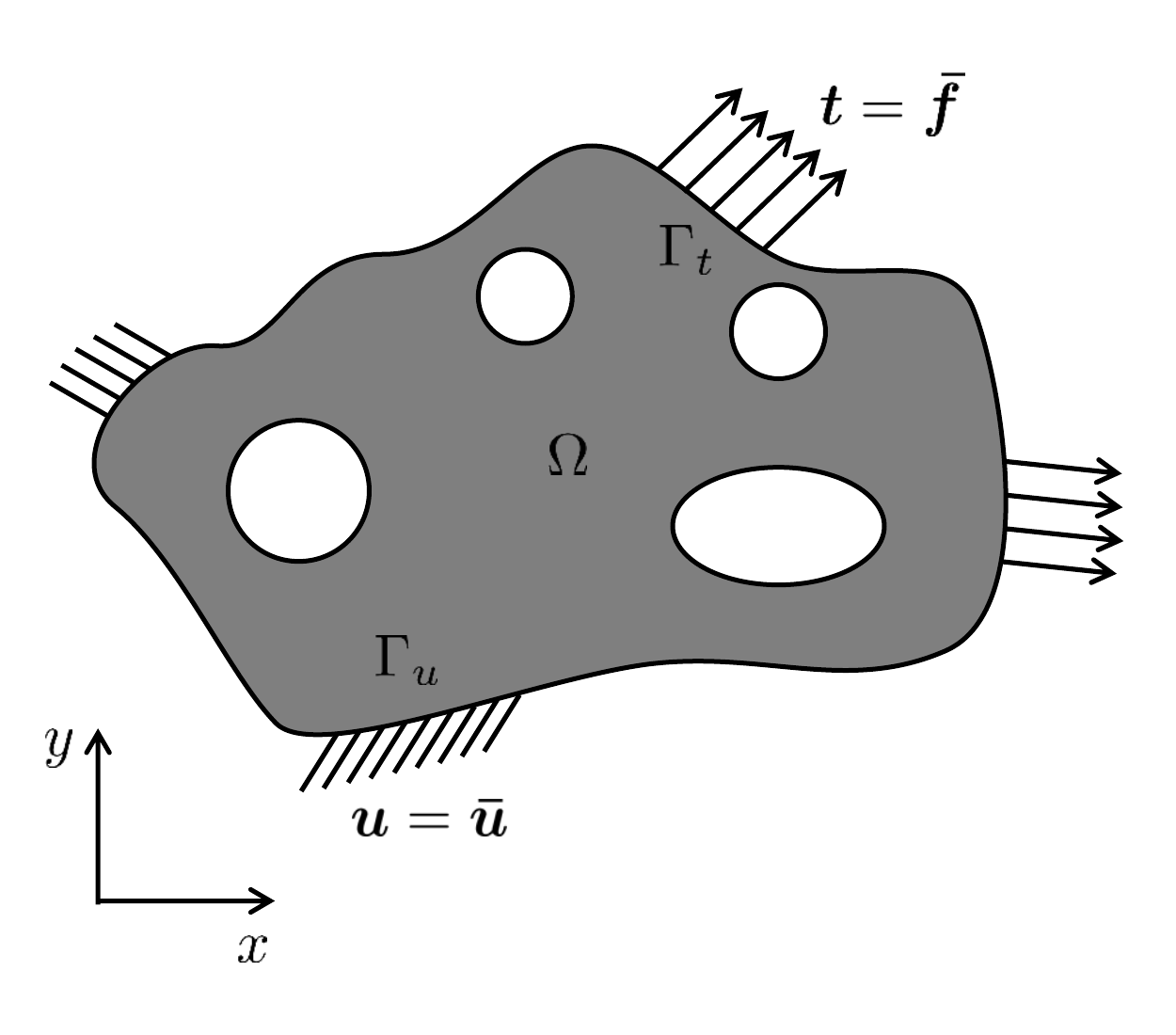}
}
\subfloat[time-dependent load profile]{
  \includegraphics[width=0.40\linewidth]{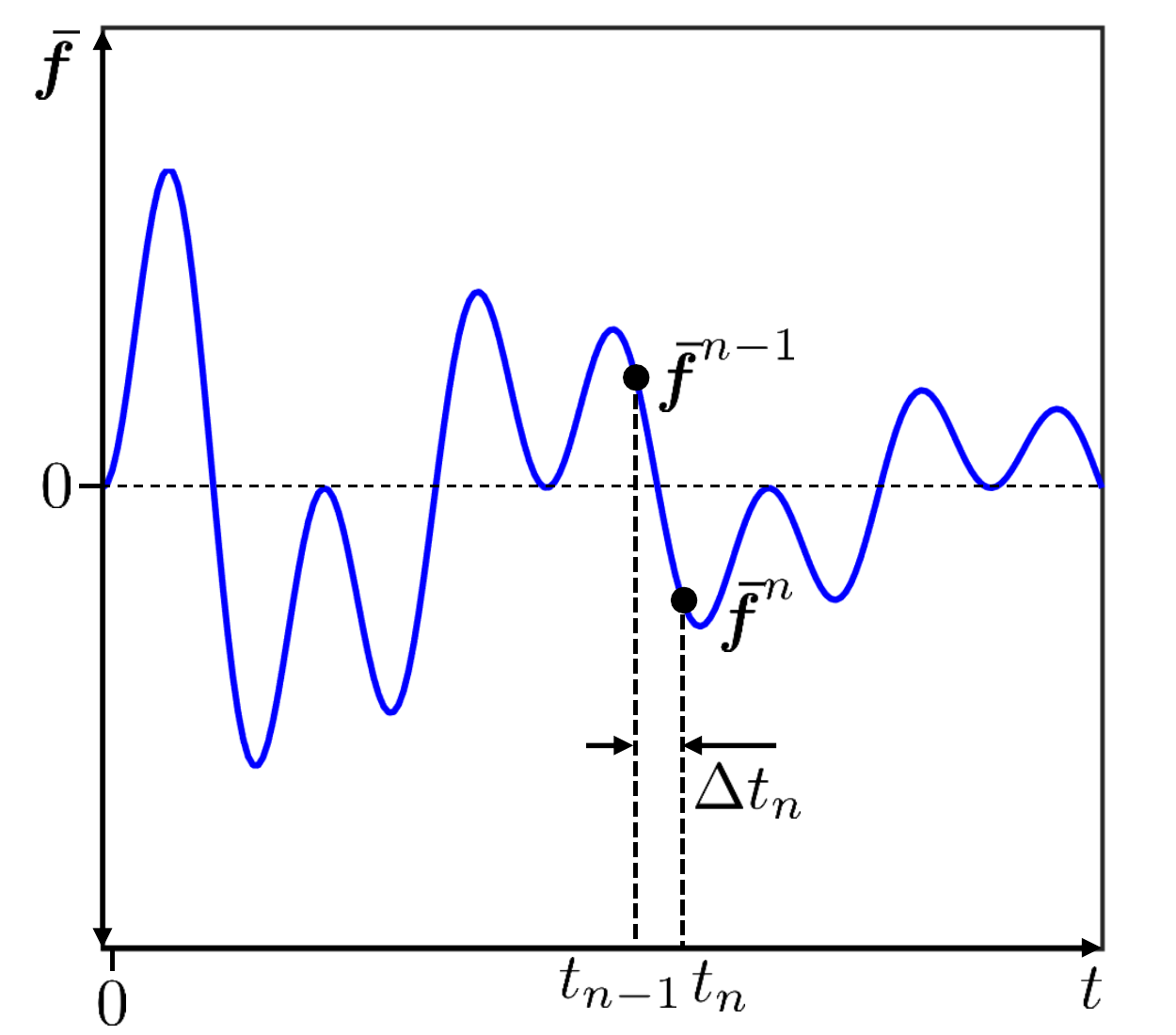}
}
\caption{Representative configuration of an optimization problem with time-dependent load profile.}
\label{fig:OptimizationProblem}
\end{figure}

\mathleft
\begin{equation*}\label{eq:DO_SensitivityAnalysis_1}
\begin{split}
\underset{\boldsymbol{\phi}}{\text{min}} \qquad \qquad Z( \boldsymbol{\rho(\boldsymbol{\phi})}, \boldsymbol{u}^{0}, ..., \,\boldsymbol{u}^{N_{t}}) = \sum_{n = 0}^{N_{t}} z^{n} ( \boldsymbol{\rho}(\boldsymbol{\phi}), \boldsymbol{u}^{n})
\end{split}
\end{equation*}
\begin{equation}\label{eq:DO_SensitivityAnalysis_2}
\text{subject to}
\begin{cases}
H_{i}( \boldsymbol{\rho(\boldsymbol{\phi})}, \boldsymbol{u}^{0}, ..., \,\boldsymbol{u}^{N_{t}}) = \displaystyle \sum_{n = 0}^{N_{t}} h^{n}_{i} ( \boldsymbol{\rho}(\boldsymbol{\phi}), \boldsymbol{u}^{n}) = 0 \qquad \forall i = 1, ..., N_{h} \\
G_{j}( \boldsymbol{\rho(\boldsymbol{\phi})}, \boldsymbol{u}^{0}, ..., \,\boldsymbol{u}^{N_{t}}) = \displaystyle \sum_{n = 0}^{N_{t}} g^{n}_{j} ( \boldsymbol{\rho}(\boldsymbol{\phi}), \boldsymbol{u}^{n}) \leq 0 \qquad \forall j = 1, ..., N_{g} \\ 
\boldsymbol{R}^{n} ( \boldsymbol{\rho(\boldsymbol{\phi})}, \boldsymbol{u}^{n}, \boldsymbol{u}^{n-1}) = \boldsymbol{0} \\
\phi_{min} \leq \phi_{k} \leq \phi_{max} \qquad \forall k = 1, ..., N_{\phi}
\end{cases}
,
\end{equation}
\mathcenter
where $Z$ is the total objective function, $\boldsymbol{\phi}$ is the vector of design variables, $\boldsymbol{\rho}$ is the vector of densities, $\boldsymbol{u}^{n}$ is the vector of state variables satisfying the discretized governing equations, $\boldsymbol{R}^{n}$, at time steps $n = 0, ..., N_{t}$. The objective function, $Z$, the equality constraint, $H_{i}$, and the inequality constraint, $G_{j}$, are functions combining performance of the structure at all or select time steps.  These functions could simply be a summation to minimize the average transient response, a p-norm function to minimize the maximum response, or some other function to achieve time-dependent goals.  The independent design variables are bounded with $\boldsymbol{\phi}_{min}$ and $\boldsymbol{\phi}_{max}$. For compactness, we collect variables at all times to column vectors as follows:
\begin{equation}\label{eq:DO_SensitivityAnalysis_3}
\begin{split}
\boldsymbol{\tilde c} = [c^{0}, ..., c^{N_{t}}]^{T}, \\
\boldsymbol{\tilde R} = [\boldsymbol{R}^{0}, ..., \boldsymbol{R}^{N_{t}}]^{T}, \\
\boldsymbol{\tilde u} = [\boldsymbol{u}^{0}, ..., \boldsymbol{u}^{N_{t}}]^{T},
\end{split}
\end{equation} 
where $c$ represents either the objective or a constraint. For a transient topology optimization problem, the computational complexities of the sensitivities dramatically increase. The time-dependent adjoint sensitives of a criterion, $\boldsymbol{\tilde c}$, the objective or a constraint, with respect to the design variable, $\phi_{k}$, can be written as:
\begin{equation}\label{eq:DO_SensitivityAnalysis_4}
\begin{split}
\frac{d \boldsymbol{\tilde c}}{d \phi_{k}} =  \frac{\partial \boldsymbol{\tilde c}}{\partial \phi_{k}} + \frac{\partial \boldsymbol{\tilde c}}{\partial \boldsymbol{\tilde u}} \frac{d \boldsymbol{\tilde u}}{d \phi_{k}}.
\end{split}
\end{equation}
The total derivative of the state variables with respect to the design variables, $d \boldsymbol{\tilde u}/d \phi_{k}$, is computed from the total derivative of the converged residual equation as follows:
\begin{equation}\label{eq:DO_SensitivityAnalysis_5}
\begin{aligned}
\frac{d \boldsymbol{\tilde R}}{d\phi_{k}} = \frac{\partial \boldsymbol{\tilde R}}{\partial \phi_{k}} + \frac{\partial \boldsymbol{\tilde R}}{\partial \boldsymbol{\tilde u}} \frac{d \boldsymbol{\tilde u}}{d \phi_{k}} = \boldsymbol{0}.
\end{aligned}
\end{equation}
Solving Equation \eqref{eq:DO_SensitivityAnalysis_5} for $d \boldsymbol{\tilde u}/d \phi_{k}$ and substituting the result into Equation \eqref{eq:DO_SensitivityAnalysis_4} yields:
\begin{equation}\label{eq:DO_SensitivityAnalysis_6}
\begin{aligned}
\frac{d \boldsymbol{\tilde c}}{d \phi_{k}} = \frac{\partial \boldsymbol{\tilde c}}{\partial \phi_{k}} + \boldsymbol{\tilde \lambda}^{T} \frac{\partial \boldsymbol{\tilde R}}{\partial \phi_{k}},
\end{aligned}
\end{equation}
where $\boldsymbol{\tilde \lambda}$, vector of adjoint solutions for all time steps, is computed by solving the following adjoint problem:
\begin{equation}\label{eq:DO_SensitivityAnalysis_7}
\begin{aligned}
\left (\frac{\partial \boldsymbol{\tilde R}}{\partial \boldsymbol{\tilde u}}  \right )^{T} \boldsymbol{\tilde \lambda} = - \left ( \frac{\partial \boldsymbol{\tilde c}}{\partial \boldsymbol{\tilde u}} \right )^{T}.
\end{aligned}
\end{equation}
For the elastodynamics problem, at the zeroth step, $n = 0$, the derivatives of the residual vector with respect to the state vectors, $\partial \boldsymbol{R}^{0} / \partial \boldsymbol{u}^{j}$, are given by:
\begin{equation}\label{eq:DO_SensitivityAnalysis_8}
\begin{aligned}
\frac{\partial \boldsymbol{R}^{0}}{\partial \boldsymbol{u}^{j}}
=
\begin{cases}
\boldsymbol{I} &  \forall j = 0 \\ 
\mathbf{0} & \forall j = 1, ..., N_{t}
\end{cases}
,
\end{aligned}
\end{equation}
where $\boldsymbol{I}$ is the identity matrix. For the $n$th step, $n>0$, the derivatives of $\partial \boldsymbol{R}^{n} / \partial \boldsymbol{u}^{j}$ yield:
\begin{equation}\label{eq:DO_SensitivityAnalysis_9}
\begin{aligned}
\frac{\partial \boldsymbol{R}^{n}}{\partial \boldsymbol{u}^{j}}
=
\begin{cases}
- \frac{1}{\tilde \beta \Delta t^{2}} \boldsymbol{M} - \frac{\tilde \gamma}{\tilde \beta \Delta t} \boldsymbol{C} &  \forall j = n - 1 \\ 
\frac{1}{\tilde \beta \Delta t^{2}} \boldsymbol{M} + \frac{\tilde \gamma}{\tilde \beta \Delta t} \boldsymbol{C} + \boldsymbol{K} &  \forall j = n \\
\mathbf{0} & \forall j \in \left \{1, ..., N_{t}\right \} \backslash \left \{n - 1, n\right \}
\end{cases}
.
\end{aligned}
\end{equation}
The derivatives of the residual equation with respect to the design variables, $\partial \boldsymbol{\tilde R} / \partial \phi_{k}$, can be computed analytically or numerically. From the residual equation, \eqref{eq:EquationofMotion_4}, the analytical derivative of the residual with respect to the design variables is described as follows:
\begin{equation}\label{eq:DO_SensitivityAnalysis_12}
\begin{split}
\frac{\partial \boldsymbol{R}^{n}}{\partial \phi_{k}} = \left [ \left( \frac{1}{2\tilde \beta} -1 \right) \frac{\partial \boldsymbol{M}}{\partial \phi_{k}} + \left( 1 - \frac{\tilde \tilde \gamma}{2\tilde \beta} \right) \Delta t \frac{\partial \boldsymbol{C}}{\partial \phi_{k}} \right ] \boldsymbol{\ddot u}^{n-1} \\
- \left [ \frac{1}{\tilde \beta \Delta t} \frac{\partial \boldsymbol{M}}{\partial \phi_{k}} - \left( 1 - \frac{\tilde \gamma}{\tilde \beta} \right) \frac{\partial \boldsymbol{C}}{\partial \phi_{k}} \right ] \boldsymbol{\dot u}^{n-1} \\
+ \left [ \frac{1}{\tilde \beta \Delta t^{2}} \frac{\partial \boldsymbol{M}}{\partial \phi_{k}} + \frac{\tilde \gamma}{\tilde \beta \Delta t} \frac{\partial \boldsymbol{C}}{\partial \phi_{k}} \right ] \left( \boldsymbol{u}^{n} - \boldsymbol{u}^{n-1} \right) + \frac{\partial \boldsymbol{K}}{\partial \phi_{k}} \boldsymbol{u}^{n}.
\end{split}
\end{equation}
For problems considered here, $\partial \boldsymbol{C}/\partial \phi_{k} = \boldsymbol{0}$. Using the SIMP approach \cite{Bendsoe:89}, the stiffness of the structure is related to the topology and the Young's modulus of an element is described as follows:
\begin{equation}\label{eq:DO_SensitivityAnalysis_13}
\begin{split}
E^{e} (\rho) = \big( (\rho^{e})^{\eta} + \rho_{min}^{e} \big) E_{0}^{e},
\end{split}
\end{equation}
where $\rho^{e}$ is the elemental density, $\eta \geq 1$ is the exponent penalty term, $E_{0}^{e}$ is the Young's modulus of a pure solid material and $\rho_{min}^{e}$ is a small positive number to maintain positive definiteness of the global stiffness matrix. In the HPM, the design variables associated with a material phase are projected onto the finite elements by a regularized Heaviside function \cite{GPB:04}. The connection between the design variable space, where the optimization is performed, and the finite element space, where the physical equilibrium is solved, is established through a radial projection. In this method, the projection radius can easily be chosen as the minimum desired feature size, $r_{min}$. The design variables are mapped onto the elements by computing the weighted average, $\mu^{e} (\phi)$, using linear filtering scheme as follows:
\begin{equation}\label{eq:DO_SensitivityAnalysis_14}
\begin{aligned}
\mu^{e}(\phi) = \frac{\displaystyle \sum_{j \in \Omega_{w}} w_{j} \phi_{j}}{\displaystyle \sum_{j \in \Omega_{w}} w_{j}}, \qquad \text{with} \qquad w_{j} = 
\begin{cases}
\frac{r_{min} - \parallel x_{j} - x^{e} \parallel}{r_{min}} & \text{ if } x \in \Omega_{w} \\ 
0 & \text{otherwise} 
\end{cases},
\end{aligned}
\end{equation}
where $x_{j}$ is the position of node $j$, and $x^{e}$ is the central position of element $e$. The elemental density is related to the design variables, $\phi$, as follows:
\begin{equation}\label{eq:DO_SensitivityAnalysis_15}
\begin{aligned}
\rho^{e} (\phi) = 1 - e^{- \beta \mu^{e} (\phi)} + \frac{\mu^{e} (\phi)}{\phi_{max}} e^{- \beta \phi_{max}},
\end{aligned}
\end{equation}
where $\beta \geq 0$ dictates the curvature of the regularization. For more information about the HPM the reader is referred to \cite{GPB:04, GAH:11}.
%

%
\section{Numerical examples
        \label{sec:NumericalExamples}}
The method is applied to a structural problem using variety of design considerations. To get insight into the performance of optimization framework, we consider both static and dynamic optimization problems. Throughout this section, a 2D cantilever beam subjected to static and dynamic loading conditions is considered. The schematic of the beam with applied load and boundary conditions is shown in Figure \ref{fig:Cantilever_Beam_Sin_Cos}, where the loads are applied at points $A$ and $B$. The model and material parameters used in the static and dynamic designs are given in Table \ref{tab:NominalDesignParameters}. We study the influence of static and dynamic loading on the optimized design through numerical examples. Two different optimization cases are considered in static and dynamic designs. In the first case, the optimization problem is solved to minimize the strain energy of the structure subjected to simultaneously applied forces $\boldsymbol{\bar f}_{1} = f_{\mathrm{max}} \sin (t) $ and $\boldsymbol{\bar f}_{2} = f_{\mathrm{max}} \cos (t)$. While the second optimization case considers minimizing the strain energy in the structure where $\boldsymbol{\bar f}_{1}$ and $\boldsymbol{\bar f}_{2}$ are applied as independent load cases. More specifically, this case considers two independent forward and sensitivity analyses associated with each $\boldsymbol{\bar f}_{1}$ and $\boldsymbol{\bar f}_{2}$. For dynamic examples, we further investigate the evolution of the design subjected to dynamic loading by considering three different formulations for the objective function. In these optimization formulations we seek to minimize: (1) sum of the strain energy over the entire transient response, (2) maximum of the strain energy in the time-history response, and (3) the strain energy at a select time. Similar to the static design, we solve two different optimization cases (simultaneously and independently applied forces) for each type of the objective formulation. 

The optimization problems are solved by using the MMA \cite{Svanberg:87}. For all optimization examples, we apply a continuation approach on the SIMP exponential penalty term \cite{GAH:11}. For numerical modeling, the design domain is discretized with 270 $\times$ 90 elements, and the physical response is predicted with the plane stress condition. The system of structural dynamic equations is solved using the modified Newton-Raphson iteration with an implicit time integration scheme developed by \cite{JE:94}.   
%

\subsection{Design subjected to static loading
        \label{sec:DesignSubjectedToStaticLoading}}
\begin{figure}[!ht]
\centering
\subfloat[the cantilever beam]{
  \includegraphics[width=0.45\linewidth]{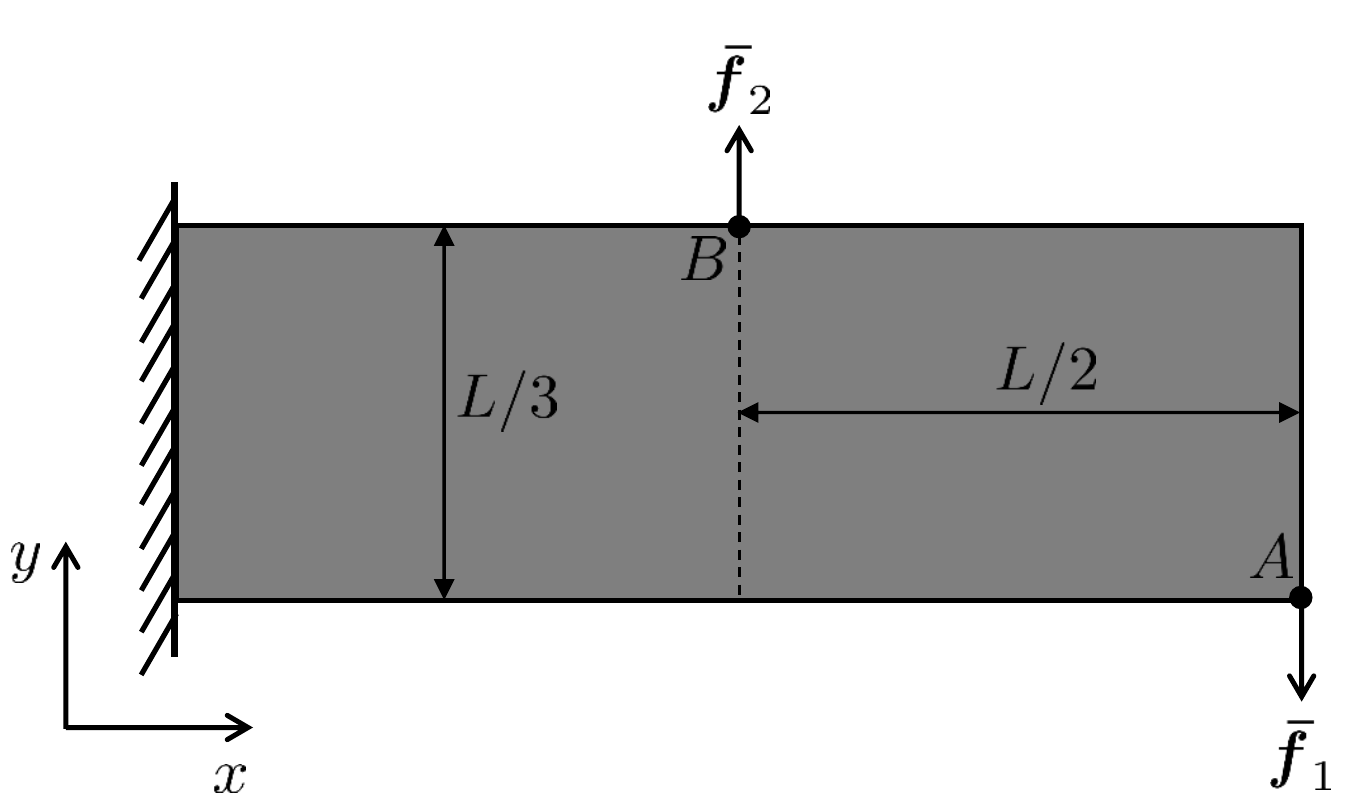}
}
\subfloat[time-dependent load profiles]{
  \includegraphics[width=0.45\linewidth]{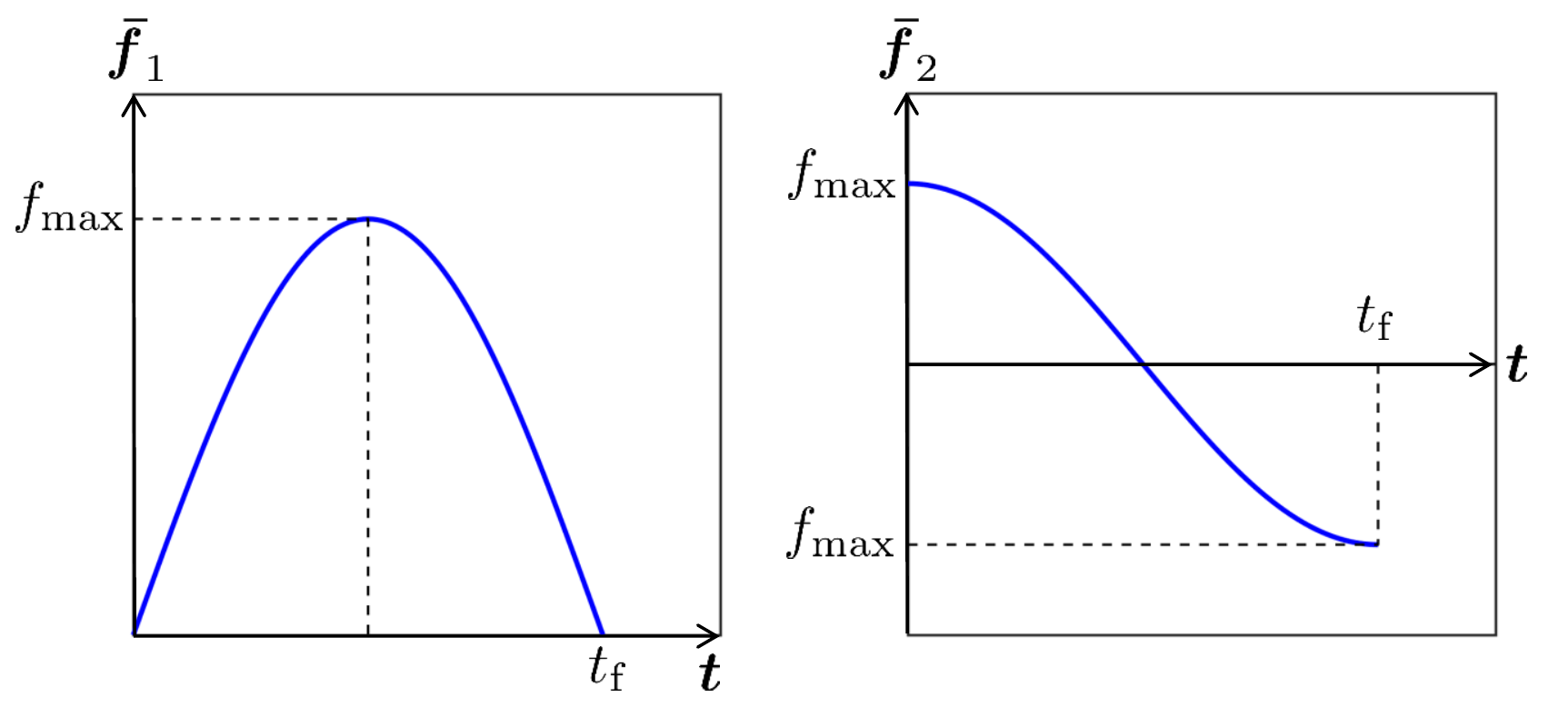}
}
\caption{Schematic of the cantilever beam with applied load and boundary conditions.}
\label{fig:Cantilever_Beam_Sin_Cos}
\end{figure}

\begin{figure}[!ht]
\centering
\subfloat[case (1): $Z = \boldsymbol{u}^{T} \boldsymbol{K} \boldsymbol{u}$]{
  \includegraphics[width=0.46\linewidth]{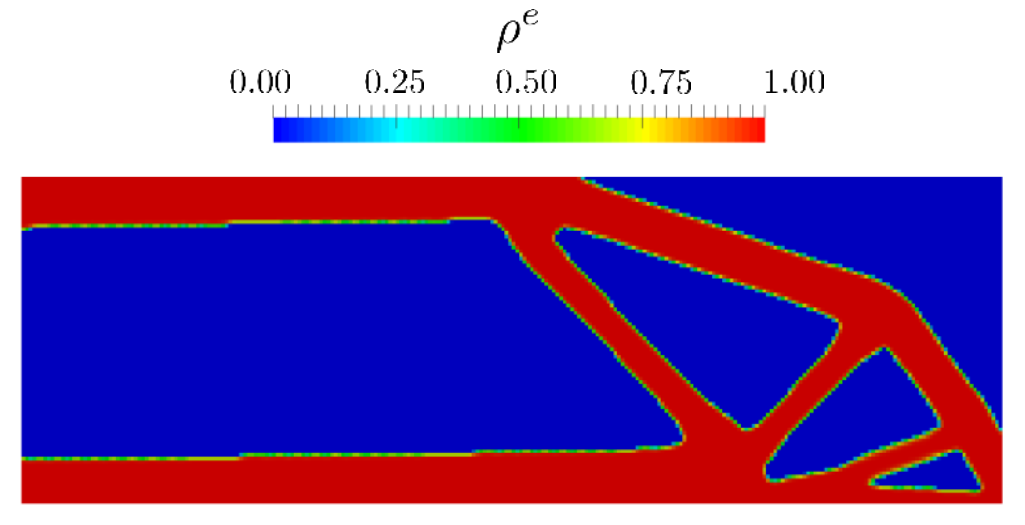}
}
\subfloat[case (2): $Z = (\boldsymbol{u}_{1})^{T} \boldsymbol{K} \boldsymbol{u}_{1} + (\boldsymbol{u}_{2})^{T} \boldsymbol{K} \boldsymbol{u}_{2}$]{
  \includegraphics[width=0.46\linewidth]{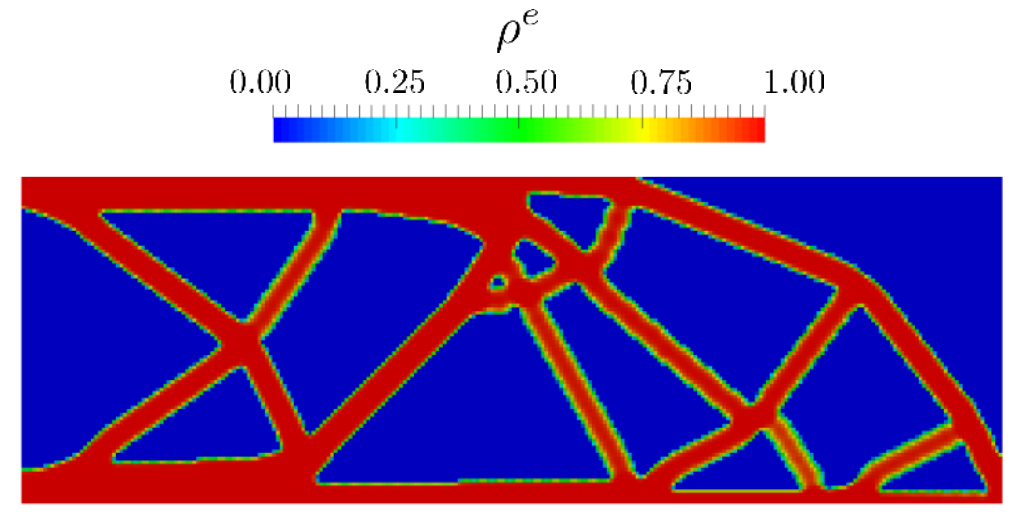}
}
\caption{Optimized structures under the static load.}
\label{fig:sin_cos_00_f1f2_1_0_0_final_design}
\end{figure}

In this section, we explore the characteristics of the proposed optimization framework through a design subjected to static loading. To this end, the 2D cantilever beam, shown in Figure \ref{fig:Cantilever_Beam_Sin_Cos}, subjected to two different loading conditions is considered. For the static design, the following optimization problems are defined. In the first optimization problem, we seek to:
\begin{equation}\label{eq:NominalDesign_1}
\begin{split}
\underset{\boldsymbol{\phi}}{\text{min}} & \quad \quad \displaystyle Z = \boldsymbol{u}^{T} \boldsymbol{K} (\phi) \boldsymbol{u} \\
\text{subject to} & \qquad \displaystyle \sum_{e \in \Omega} \rho^{e} (\phi) v^{e} \leq V_{max} \\
 & \qquad \displaystyle  \boldsymbol{K} (\phi) \boldsymbol{u} = f_{\mathrm{max}} \Big|_{A} + f_{\mathrm{max}} \Big|_{B} \\
 & \qquad \displaystyle \phi_{min} \leq \boldsymbol{\phi} \leq \phi_{max}
\end{split}
,
\end{equation}
where $\boldsymbol{u}$ is the solution corresponding to the structural response subjected to $f_{\mathrm{max}}$ applied at point $A$ and $B$, $v^{e}$ is the elemental volume, and $V_{max}$ is the maximum allowable volume fraction. In the second case, the objective is defined as sum of the strain energy corresponding to sum of the responses of two structures subjected to $f_{\mathrm{max}}$ at point $A$ and $B$, respectively:
\begin{equation}\label{eq:NominalDesign_2}
\begin{split}
\underset{\boldsymbol{\phi}}{\text{min}} & \quad \quad \displaystyle Z = (\boldsymbol{u}_{1})^{T} \boldsymbol{K} (\phi) \boldsymbol{u}_{1} + (\boldsymbol{u}_{2})^{T} \boldsymbol{K} (\phi) \boldsymbol{u}_{2} \\
\text{subject to} & \qquad \displaystyle \sum_{e \in \Omega} \rho^{e} (\phi) v^{e} \leq V_{max} \\
 & \qquad \displaystyle  \boldsymbol{K} (\phi) \boldsymbol{u}_{1} = f_{\mathrm{max}} \Big|_{A} \\
 & \qquad \displaystyle  \boldsymbol{K} (\phi) \boldsymbol{u}_{2} = f_{\mathrm{max}} \Big|_{B} \\
  & \qquad \displaystyle \phi_{min} \leq \boldsymbol{\phi} \leq \phi_{max}
\end{split}
,
\end{equation}
where $\boldsymbol{u}_{1}$ and $\boldsymbol{u}_{2}$ are the solutions corresponding to structural responses subjected to $f_{\mathrm{max}}$ at point $A$ and $B$, respectively. The optimized structures for the static designs are shown in Figure \ref{fig:sin_cos_00_f1f2_1_0_0_final_design}. As expected, the optimized structures are different for two optimization problems. Interestingly, in the first case where the structure is subjected to simultaneously applied loads, the shear stresses are zero from the root to midspan (point $B$) leading to the absence of internal bracing, whereas for case (2) the shear stresses are non-zero in this region for both independently applied loads.
\begin{table}[t]
\centering
\caption{Model and material parameters for the cantilever beam model.}
\label{tab:NominalDesignParameters}
\begin{tabular}{lccc}
\hline
Description & Parameter & Value & Units \\ \hline
minimum allowable radius & $r_{min}$ & $1.375 \times 10^{-2}$ & m \\
elemental length & $h_{l}$ & $3.333 \times 10^{-3}$ & m \\
minimum allowable density & $\rho_{min}$ & $1.0 \times 10^{-4}$ &  \\
Young's modulus of pure solid & $E_{0}$ & $2.0 \times 10^{11}$ & Pa \\
Poisson's ratio & $\nu$ & 0.33 &  \\
beam length & $L$ & 0.9 & m \\
beam thickness & $t_{b}$ & 0.01 & m \\
maximum applied force & $f_{\mathrm{max}}$ & $1.0 \times 10^{4}$ & N \\
maximum allowable time & $t_{\mathrm{f}}$ & $\pi$ & s \\
time step size & $\Delta t$ & $\pi/20$ & s \\
Heaviside projection parameter & $\beta$ & 50 &  \\
SIMP exponential penalty term & $\eta$ & 1,...,10 & \\
lower bound of the design variables & $\phi_{min}$ & 0.0 & \\
upper bound of the design variables & $\phi_{max}$ & 1.0 & \\
maximum allowable volume fraction & $V_{max}$ & 0.35 &  \\
\hline
\end{tabular}
\end{table}
%

\subsection{Design subjected to dynamic loading
        \label{sec:DesignWithDynamicLoading}}
For the structure subjected to dynamic loading, we consider three different type of objective functions. Similar to the static design, two different cases are considered for each objective function. In the first case, we solve the optimization problem for the structure subjected to simultaneously applied force $\boldsymbol{\bar f}_{1}$ and $\boldsymbol{\bar f}_{2}$. While in the second case, the optimization problem is solved for the structure subjected to two independent forces $\boldsymbol{\bar f}_{1}$ and $\boldsymbol{\bar f}_{2}$. The optimization problems are formulated as follows: 

case (1):
\begin{equation}\label{eq:MinimizingTotalStrainEnergy_00}
\begin{split}
\underset{\boldsymbol{\phi}}{\text{min}} & \quad \quad \displaystyle Z = \tilde Z_{1} \\
\text{subject to} & \qquad \displaystyle \sum_{e \in \Omega} \rho^{e} (\phi) v^{e} \leq V_{max} \\
 & \qquad \displaystyle  \boldsymbol{R} (\boldsymbol{\phi}, \boldsymbol{u}) = \boldsymbol{0}  \quad \text{with} \quad  \boldsymbol{\bar f} = \boldsymbol{\bar f}_{1} + \boldsymbol{\bar f}_{2} \\
   & \qquad \displaystyle \phi_{min} \leq \boldsymbol{\phi} \leq \phi_{max}
\end{split}
,
\end{equation}

case (2):
\begin{equation}\label{eq:MinimizingTotalStrainEnergy_01}
\begin{split}
\underset{\boldsymbol{\phi}}{\text{min}} & \quad \quad \displaystyle Z = \tilde Z_{2} \\
\text{subject to} & \qquad \displaystyle \sum_{e \in \Omega} \rho^{e} (\phi) v^{e} \leq V_{max} \\
 & \qquad \displaystyle  \boldsymbol{R} (\boldsymbol{\phi}, \boldsymbol{u}_{1}) = \boldsymbol{0}  \quad \text{with} \quad  \boldsymbol{\bar f} = \boldsymbol{\bar f}_{1} \\
 & \qquad \displaystyle  \boldsymbol{R} (\boldsymbol{\phi}, \boldsymbol{u}_{2}) = \boldsymbol{0}  \quad \text{with} \quad  \boldsymbol{\bar f} = \boldsymbol{\bar f}_{2} \\
 & \qquad \displaystyle \phi_{min} \leq \boldsymbol{\phi} \leq \phi_{max}
\end{split}
,
\end{equation}
where $\tilde Z_{1}$ and $\tilde Z_{2}$ represent the objective functions. The characteristics of designs subjected to dynamic loading with different objective functions are explored here.
%

\subsubsection{Minimizing the total strain energy
        \label{sec:MinimizingTotalStrainEnergy}}
\begin{figure}[!ht]
\centering
\subfloat[case (1), $Z = \tilde Z_{1}$]{
  \includegraphics[width=0.46\linewidth]{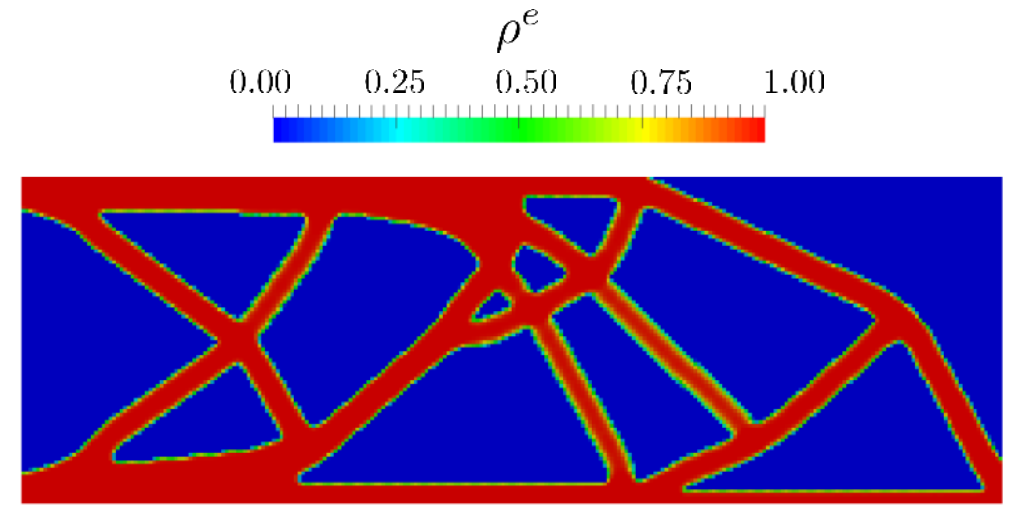}
}
\subfloat[case (2), $Z = \tilde Z_{2}$]{
  \includegraphics[width=0.46\linewidth]{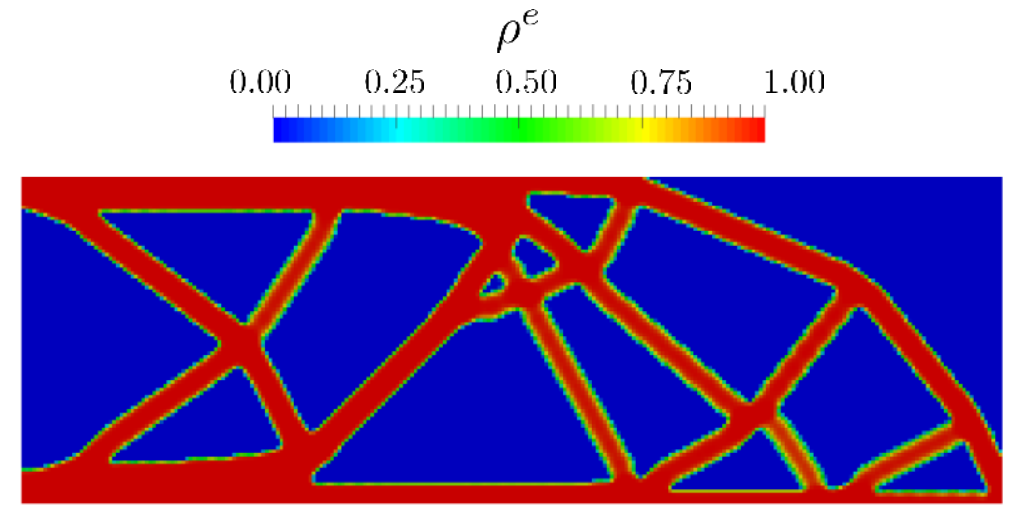}
}
\newline
\subfloat[case (1), $Z = \tilde Z_{1}$]{
  \includegraphics[width=0.46\linewidth]{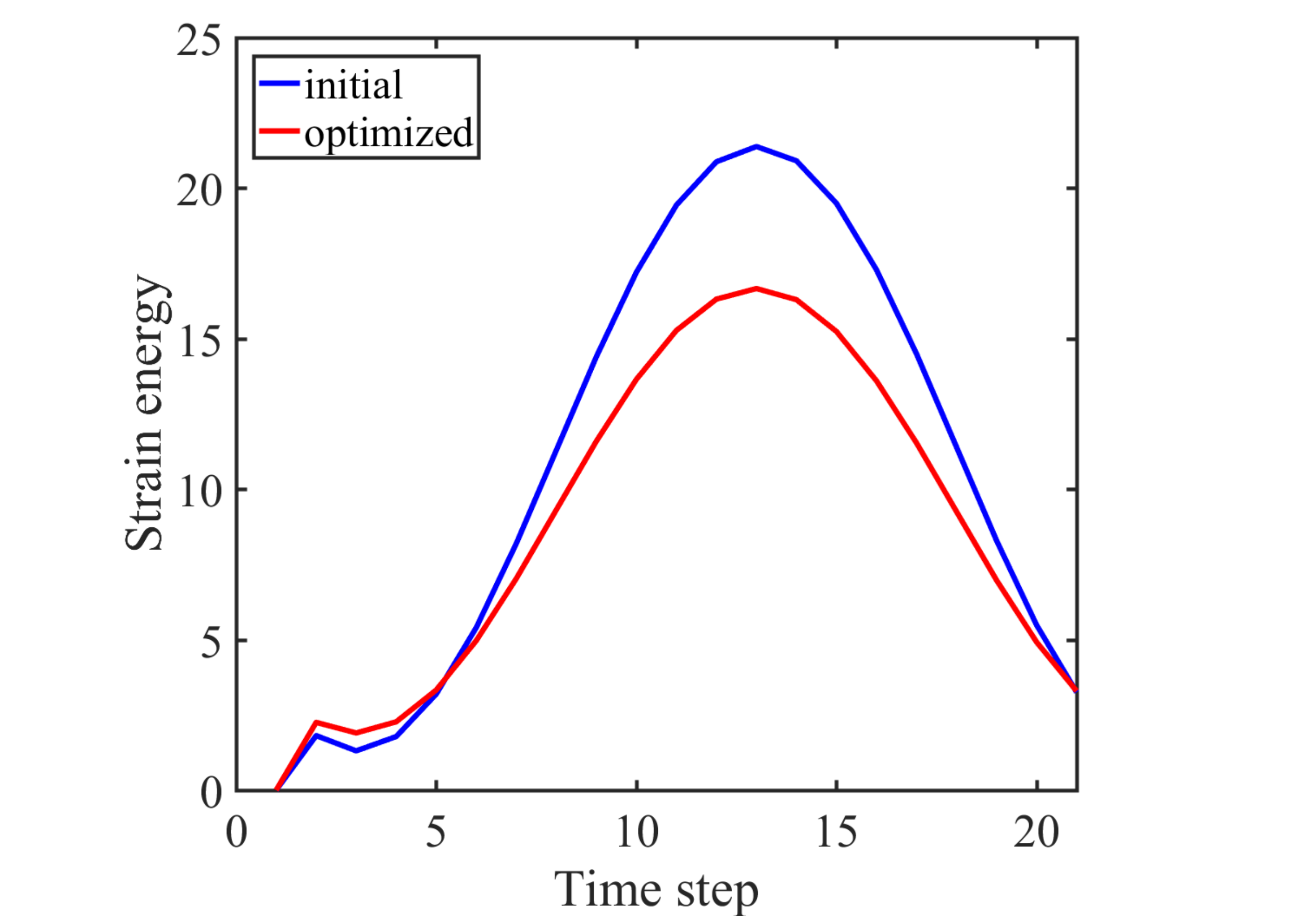}
}
\subfloat[case (2), $Z = \tilde Z_{2}$]{
  \includegraphics[width=0.46\linewidth]{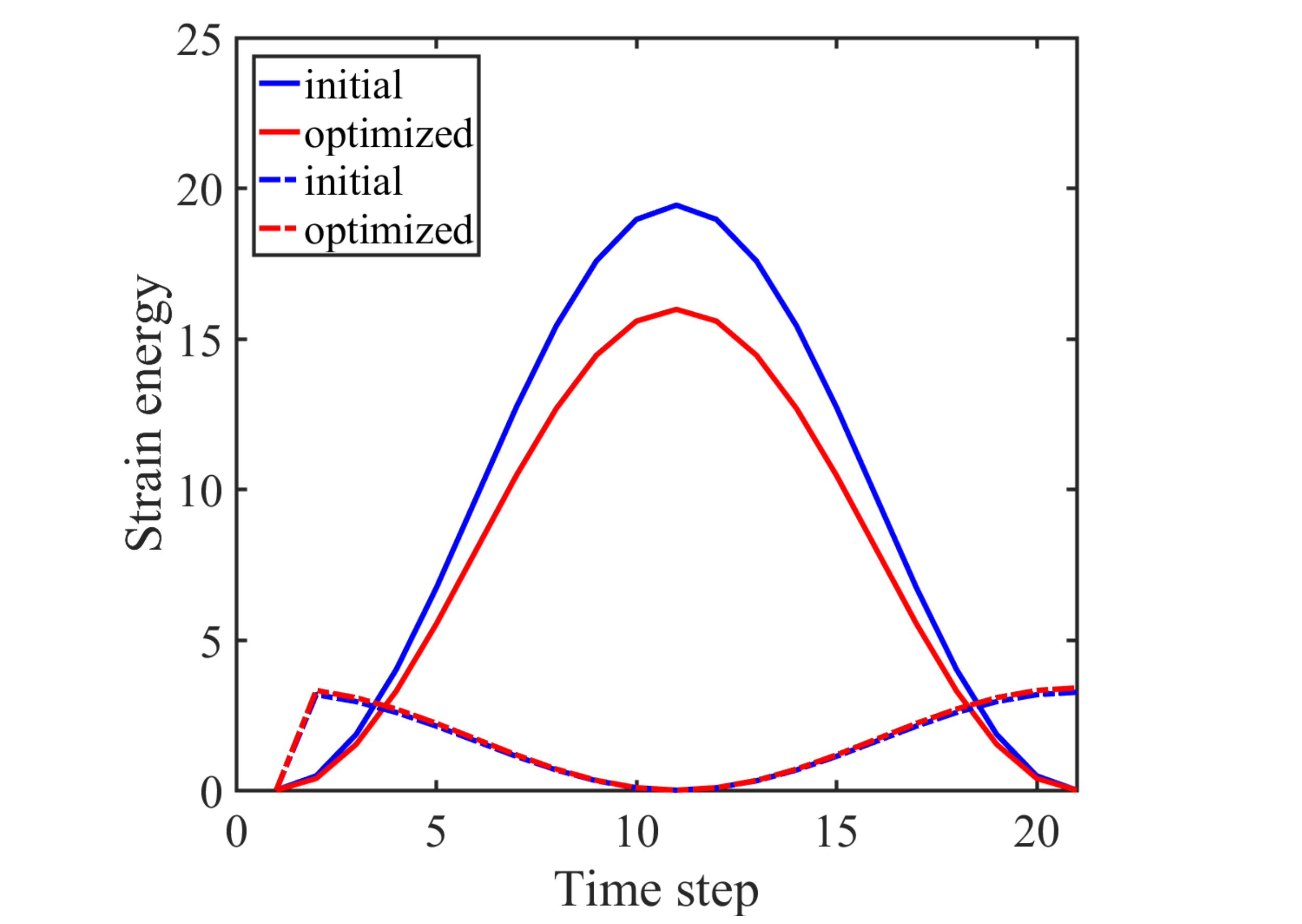}
}
\caption{Optimized structures and the strain energy profiles in the dynamic optimization problem where the total strain energy is minimized. The solid lines in (c) represent the initial and optimized strain energy profiles for the structure subjected to $\boldsymbol{\bar f}_{1} + \boldsymbol{\bar f}_{2}$, where in (d) the solid lines show the initial and optimized strain energy profiles for the structure subjected to $\boldsymbol{\bar f}_{1}$ and the dashed lines show the initial and optimized strain energy profiles for the structure subjected to $\boldsymbol{\bar f}_{2}$.}
\label{fig:sin_cos_sm_f1f2_1_0_0_final_design}
\end{figure}

For this optimization problem, we seek to minimize the sum of strain energy over the entire transient response of the structure. Similar to the static design, $\tilde Z_{1}$ and $\tilde Z_{2}$ are defined as:
\begin{equation}\label{eq:MinimizingTotalStrainEnergy_1}
\begin{split}
\tilde Z_{1} = \sum_{n = 0}^{N_{t}} (\boldsymbol{u}^{n})^{T} \boldsymbol{K} (\phi) \boldsymbol{u}^{n},
\end{split}
\end{equation}
\begin{equation}\label{eq:MinimizingTotalStrainEnergy_2}
\begin{split}
\tilde Z_{2} = \displaystyle \sum_{n = 0}^{N_{t}} (\boldsymbol{u}^{n}_{1})^{T} \boldsymbol{K} (\phi) \boldsymbol{u}^{n}_{1} + \displaystyle \sum_{n = 0}^{N_{t}} (\boldsymbol{u}^{n}_{2})^{T} \boldsymbol{K} (\phi) \boldsymbol{u}^{n}_{2},
\end{split}
\end{equation}
where $\boldsymbol{u}^{n}$, $\boldsymbol{u}^{n}_{1}$, and $\boldsymbol{u}^{n}_{2}$ are the solution vectors at time step $n$ corresponding to $\boldsymbol{\bar f}_{1} + \boldsymbol{\bar f}_{2}$, $\boldsymbol{\bar f}_{1}$, and $\boldsymbol{\bar f}_{2}$, respectively. The optimized structures are shown in Figure \ref{fig:sin_cos_sm_f1f2_1_0_0_final_design}. The results show two different designs for case (1) and (2). Moreover, both designs differ from the static designs. This is related to the dynamic response of the structure and type of the objective function that considers the structural response at all time steps. Comparison of the optimized strain energy profiles for two cases shows that in case (2), the strain energy associated with the response of structure under $\boldsymbol{\bar f}_{2}$ is minor and the design evolves to minimize the governing strain energy.   
%

\subsubsection{Minimizing the maximum of strain energy
        \label{sec:MinimizingMaxStrainEnergy}}
\begin{figure}[!ht]
\centering
\subfloat[case (1), $Z = \tilde Z_{1}$]{
  \includegraphics[width=0.46\linewidth]{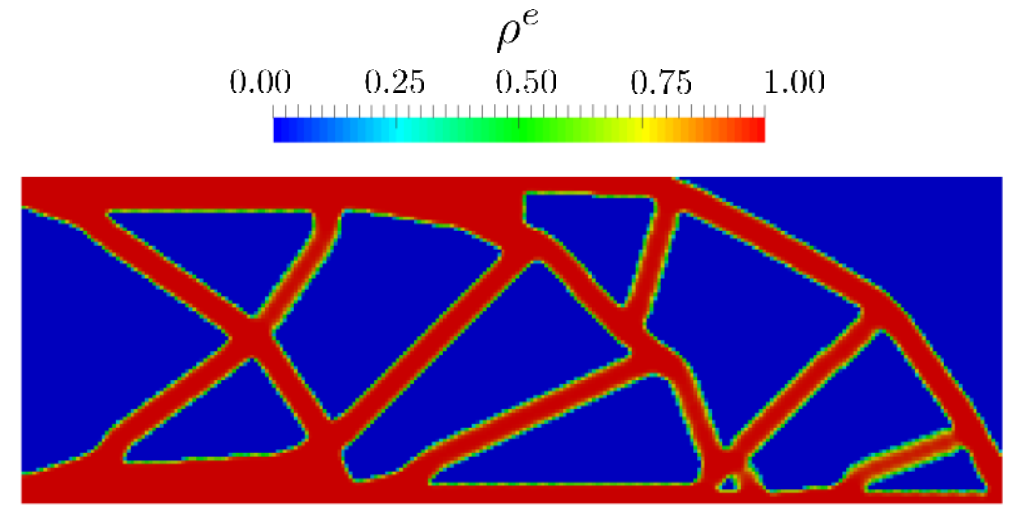}
}
\subfloat[case (2), $Z = \tilde Z_{2}$]{
  \includegraphics[width=0.46\linewidth]{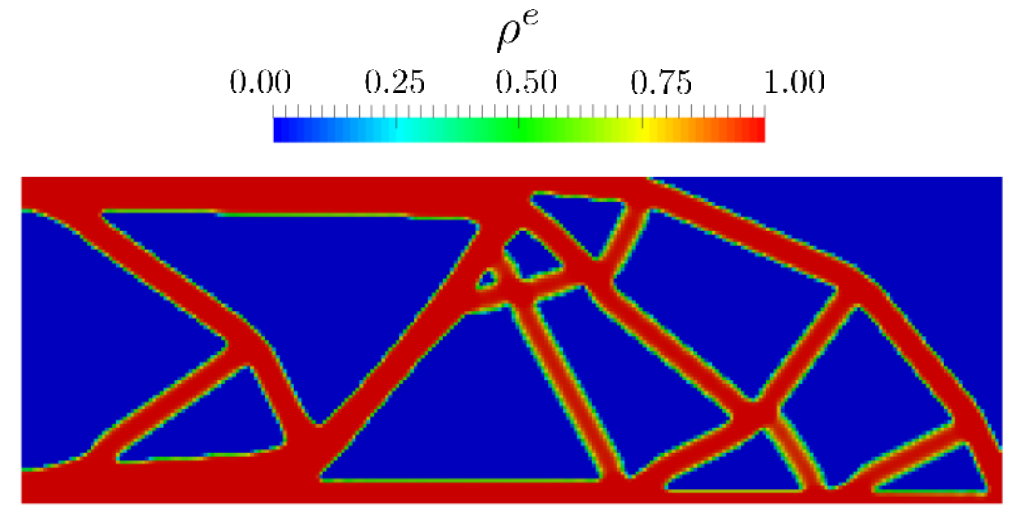}
}
\newline
\subfloat[case (1), $Z = \tilde Z_{1}$]{
  \includegraphics[width=0.46\linewidth]{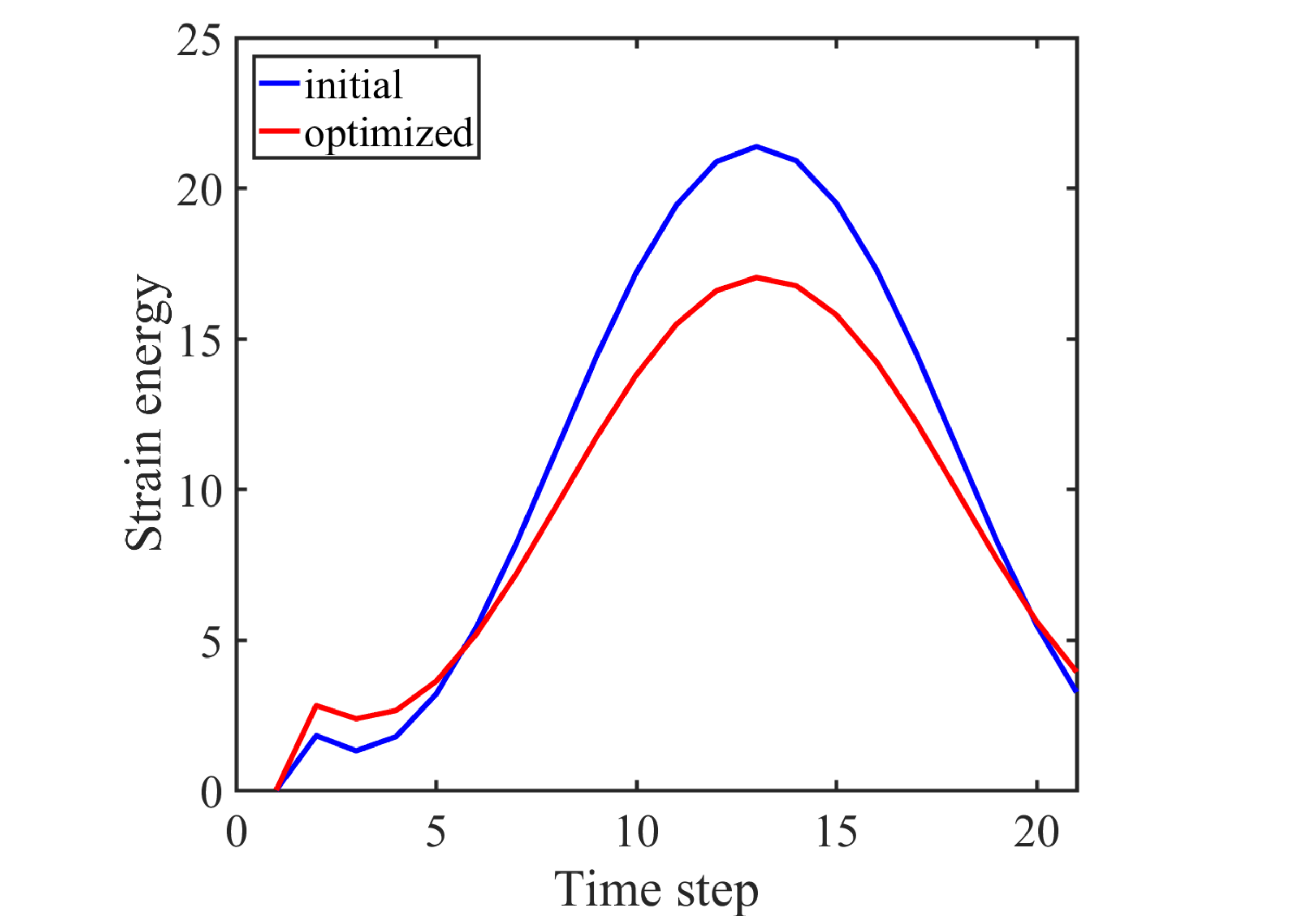}
}
\subfloat[case (2), $Z = \tilde Z_{2}$]{
  \includegraphics[width=0.46\linewidth]{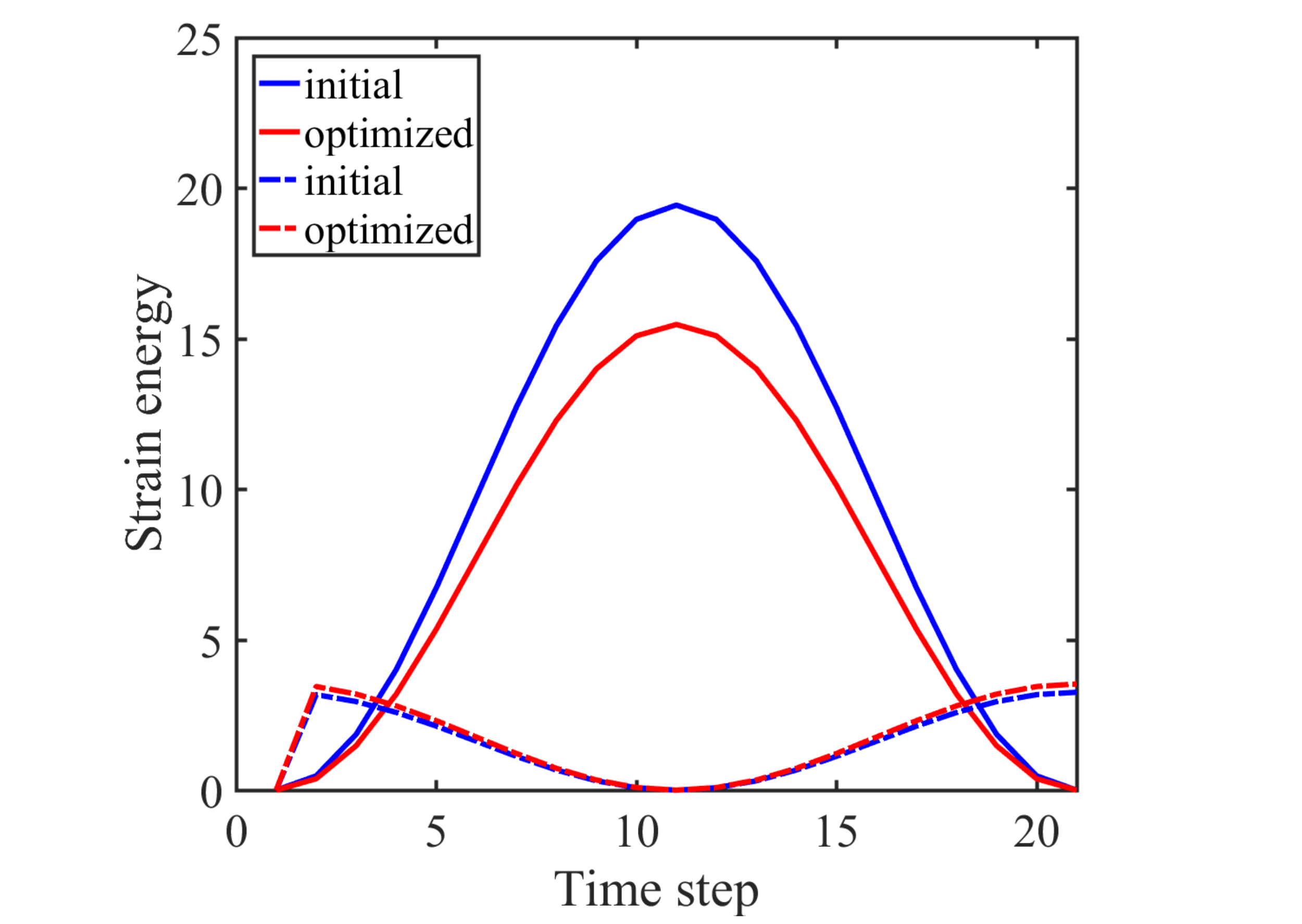}
}
\caption{Optimized structures and strain energy profiles in the dynamic optimization problem where the maximum of strain energy is minimized. The solid lines in (c) represent the initial and optimized strain energy profiles for the structure subjected to $\boldsymbol{\bar f}_{1} + \boldsymbol{\bar f}_{2}$, where in (d) the solid lines show the initial and optimized strain energy profiles for the structure subjected to $\boldsymbol{\bar f}_{1}$ and the dashed lines show the initial and optimized strain energy profiles for the structure subjected to $\boldsymbol{\bar f}_{2}$.}
\label{fig:sin_cos_ks_f1f2_1_0_0_final_design}
\end{figure}

For this case, we wish to minimize the maximum value of the strain energy over the entire response history of the structure. To construct a differentiable objective function for a min-max optimization problem, we employ the Kreisselmeier-Steinhauser (KS) formulation \cite{Kreisselmeier:79} and formulate the objective functions in the following settings:
\begin{equation}\label{eq:MinimizingMaxStrainEnergy_1}
\begin{split}
\tilde Z_{1} = \frac{1}{\bar \xi} \log_{e} \displaystyle \sum_{n = 0}^{N_{t}} \exp \big ( \bar \xi \frac{(\boldsymbol{u}^{n})^{T} \boldsymbol{K} (\phi) \boldsymbol{u}^{n}}{z^{0}_{\mathrm{max}}} \big ),
\end{split}
\end{equation}
\begin{equation}\label{eq:MinimizingMaxStrainEnergy_2}
\begin{split}
\tilde Z_{2} = \displaystyle \frac{1}{\bar \xi} \log_{e} \sum_{n = 0}^{N_{t}} \exp \big ( \bar \xi \frac{(\boldsymbol{u}^{n}_{1})^{T} \boldsymbol{K} (\phi) \boldsymbol{u}^{n}_{1}}{z^{0}_{\mathrm{max}}} \big ) + \displaystyle \frac{1}{\bar \xi} \log_{e} \sum_{n = 0}^{N_{t}} \exp \big ( \bar \xi \frac{(\boldsymbol{u}^{n}_{2})^{T} \boldsymbol{K} (\phi) \boldsymbol{u}^{n}_{2}}{z^{0}_{\mathrm{max}}} \big ),
\end{split}
\end{equation}
where $\bar \xi$ is a scalar parameter in the KS function which caries between 5 and 200 \ \cite{Kreisselmeier:79} and $z^{0}_{\mathrm{max}}$ is the maximum value of the strain energy in the time history of the initial design. In this study we set $\bar \xi = 20$. The optimized designs are shown in Figure \ref{fig:sin_cos_ks_f1f2_1_0_0_final_design}. The trend of designs is similar to the designs presented in Section \ref{sec:MinimizingTotalStrainEnergy}; however, the optimized structures are different in case (1) and (2). Moreover, similar to case (2) in Section \ref{sec:MinimizingTotalStrainEnergy}, the strain energy associated with the response of structure under $\boldsymbol{\bar f}_{2}$ is negligible during the optimization. These observations indicate the importance in the timing, location, and magnitude of the applied load for the structural dynamics.
%

\subsubsection{Minimizing strain energy at a designer selected time
        \label{sec:MinimizingStrainEnergyAtTime}}
\begin{figure}[!ht]
\centering
\subfloat[case 1]{
  \includegraphics[width=0.46\linewidth]{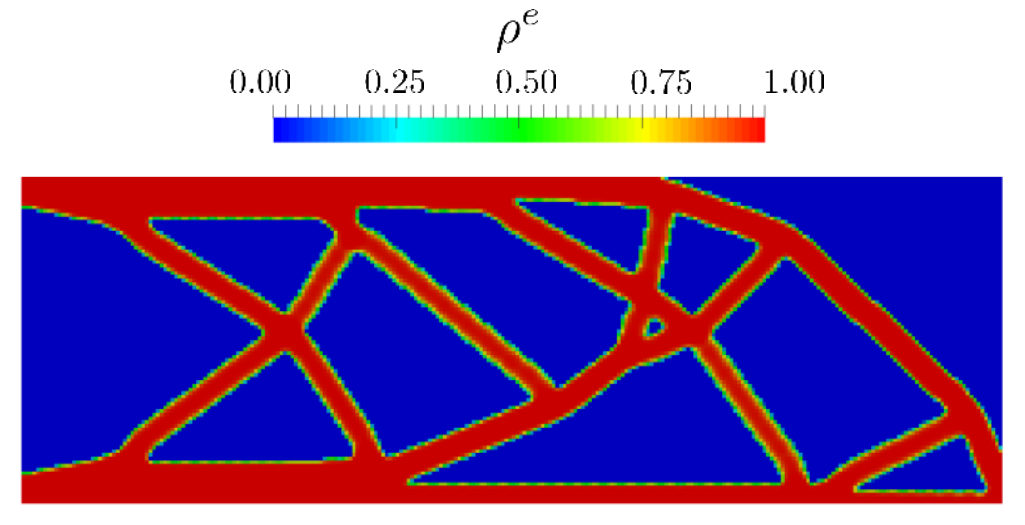}
}
\subfloat[case 2]{
  \includegraphics[width=0.46\linewidth]{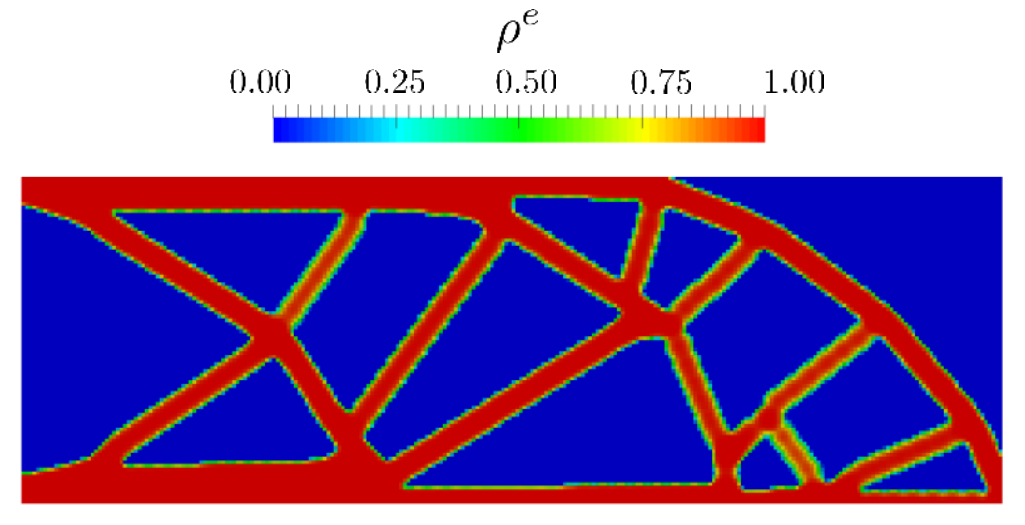}
}
\newline
\subfloat[case 1]{
  \includegraphics[width=0.46\linewidth]{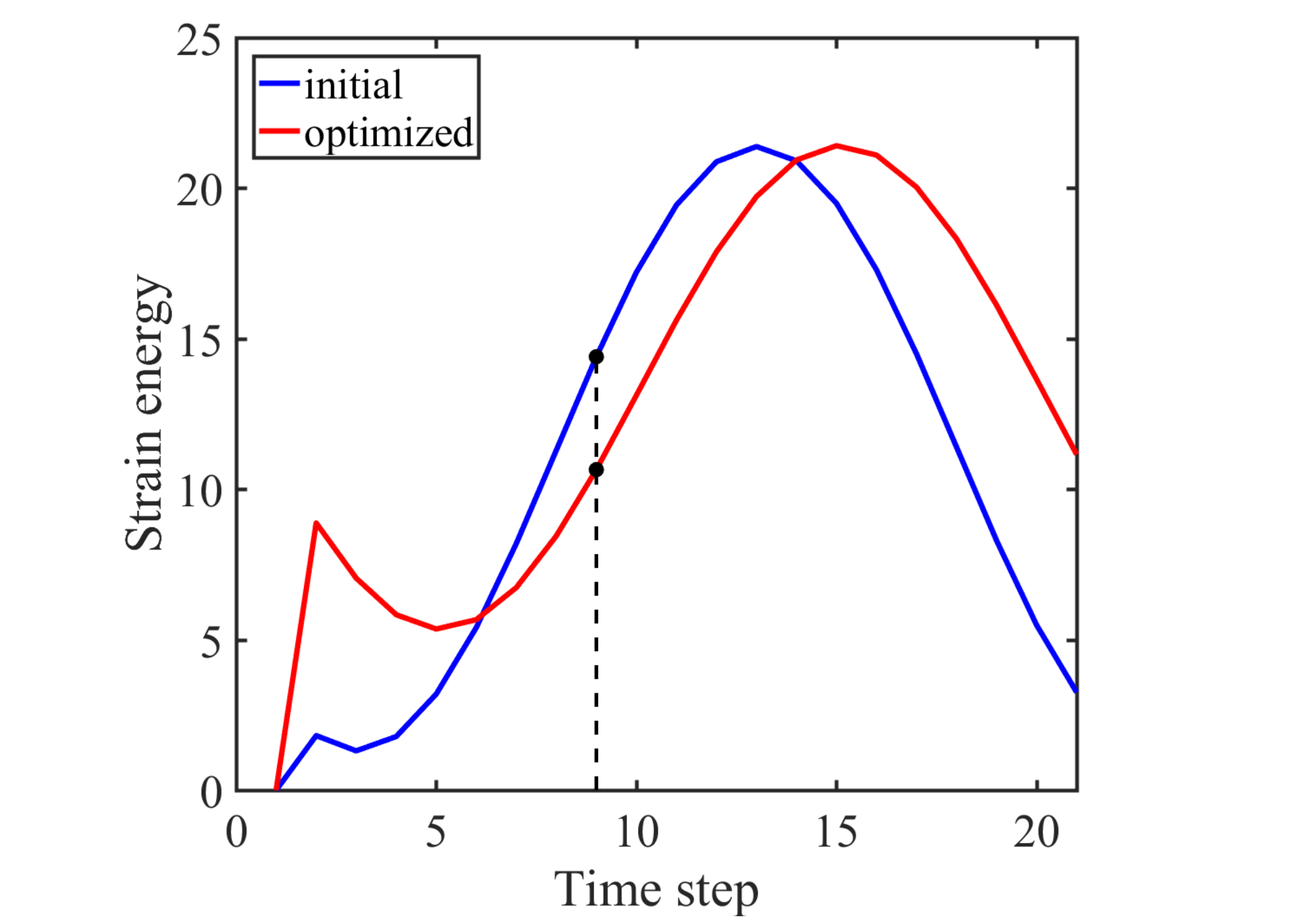}
}
\subfloat[case 2]{
  \includegraphics[width=0.46\linewidth]{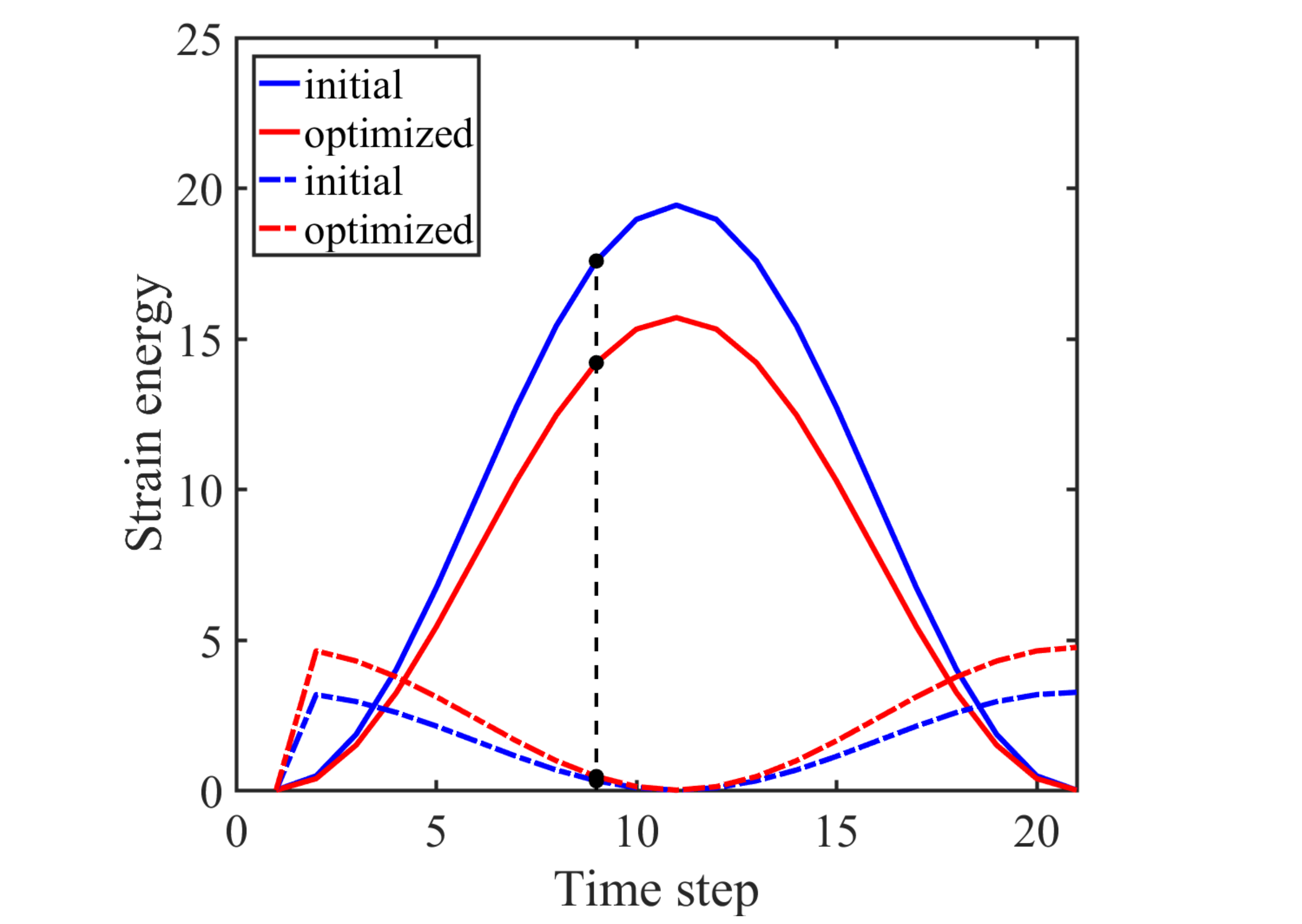}
}
\caption{Optimized structures and strain energy profiles in the dynamic optimization problem where the strain energy at a given time is minimized. The solid lines in (c) represent the initial and optimized strain energy profiles for the structure subjected to $\boldsymbol{\bar f}_{1} + \boldsymbol{\bar f}_{2}$, while in (d) the solid lines show the initial and optimized strain energy profiles for the structure subjected to $\boldsymbol{\bar f}_{1}$ and the dashed lines show the initial and optimized strain energy profiles for the structure subjected to $\boldsymbol{\bar f}_{2}$.}
\label{fig:sin_cos_att_f1f2_1_0_0_final_design}
\end{figure}

\begin{figure}[!ht]
\centering
\subfloat[case (1)]{
  \includegraphics[width=0.46\linewidth]{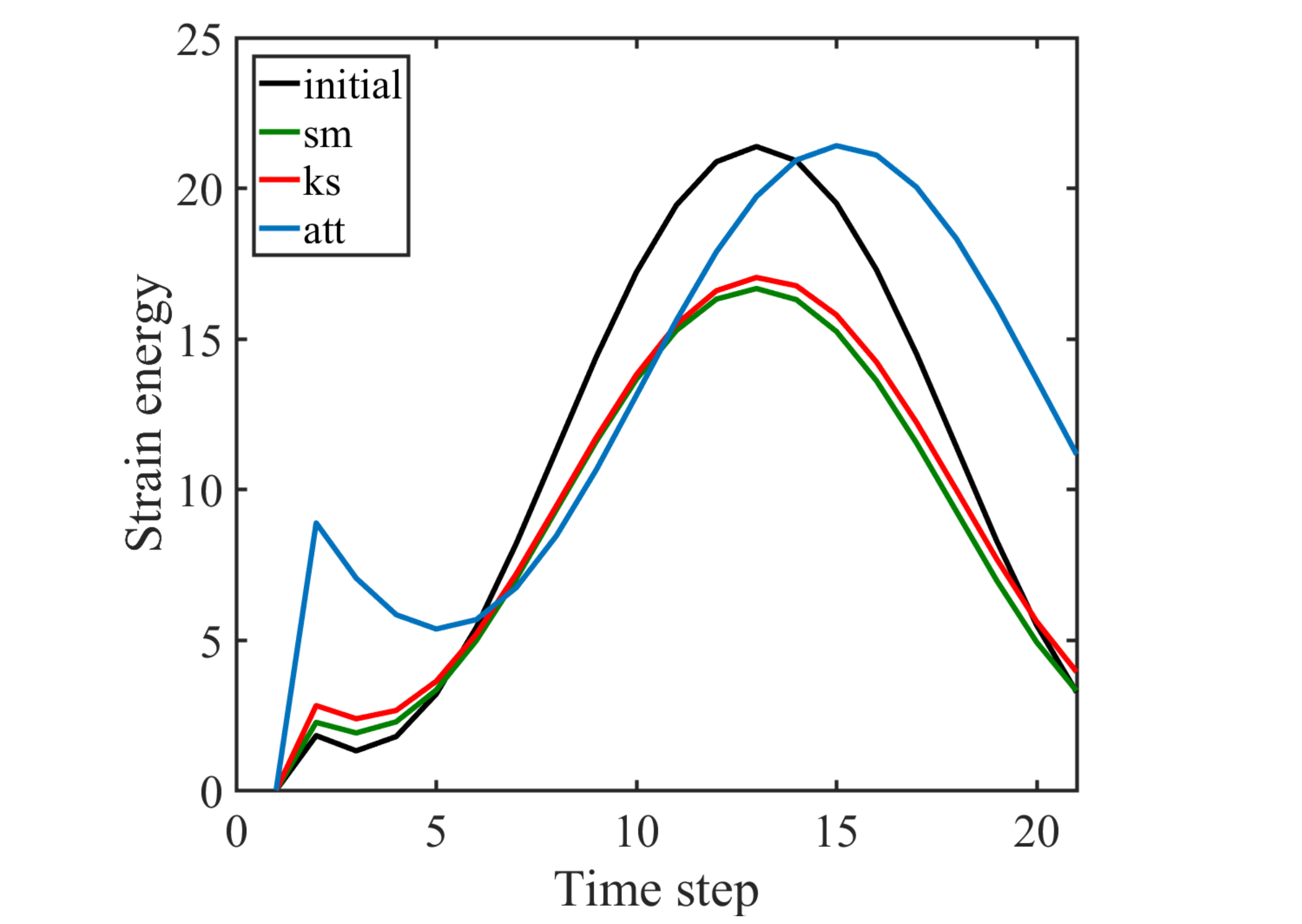}
}
\subfloat[case (2)]{
  \includegraphics[width=0.46\linewidth]{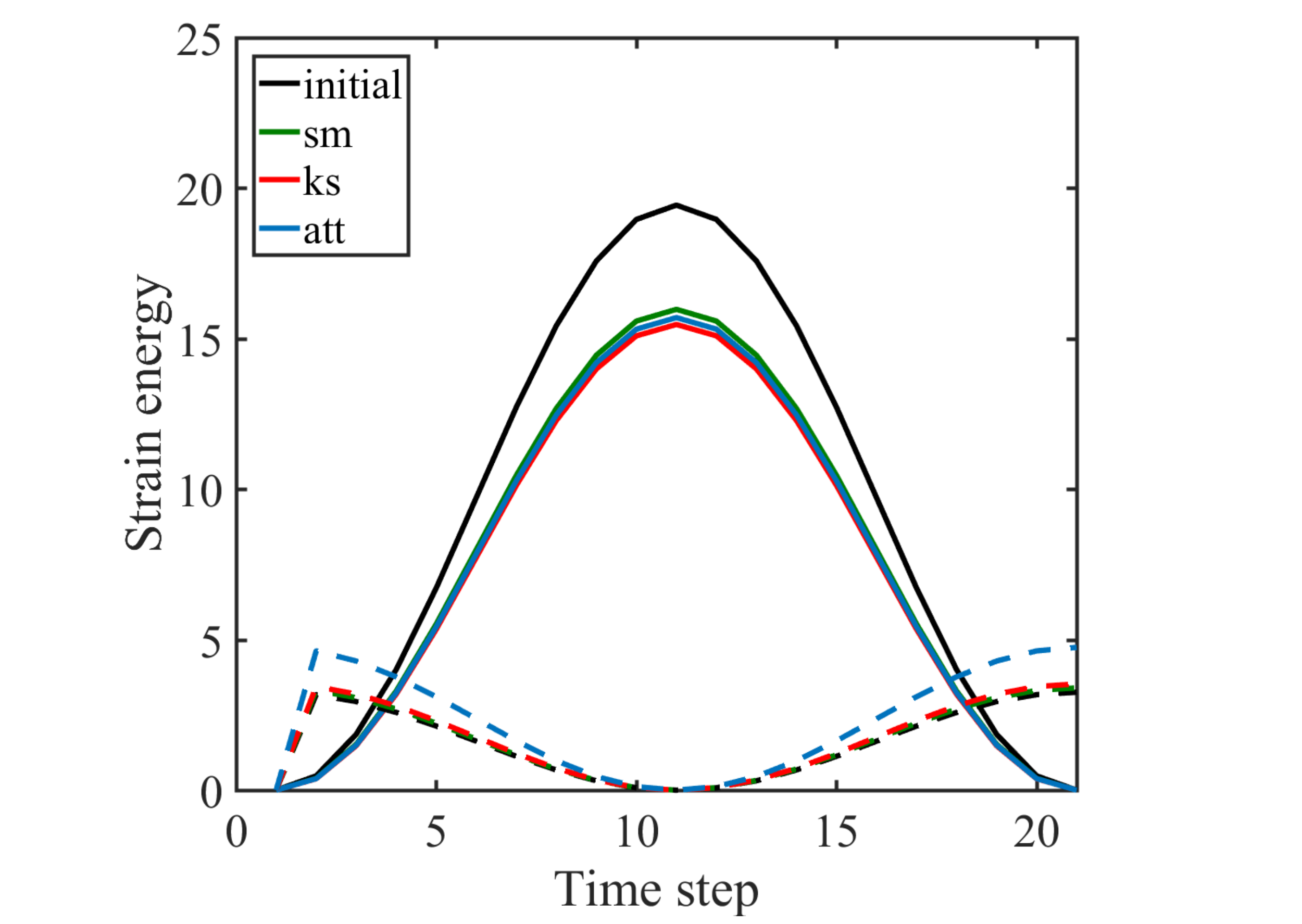}
}
\caption{Comparison of the strain energy profile for the optimized designs under dynamic loading. The objective formulations, \textit{sm}: minimizing the total strain energy, \textit{ks}: minimizing the maximum of the strain energy, and \textit{att}: minimizing the strain energy at a select time.}
\label{fig:compar_strain_energy_sm_ks_att}
\end{figure}

Finally, in this section we seek to minimize the strain energy at a time in the time history profile that is selected by the designer. To this end, we define the following objective functions:
\begin{equation}\label{eq:MinimizingStrainEnergyAtTime_1}
\begin{split}
\tilde Z_{1} = (\boldsymbol{u}^{t})^{T} \boldsymbol{K} (\phi) \boldsymbol{u}^{t},
\end{split}
\end{equation}
\begin{equation}\label{eq:MinimizingStrainEnergyAtTime_2}
\begin{split}
\tilde Z_{2} = (\boldsymbol{u}_{1}^{t})^{T} \boldsymbol{K} (\phi) \boldsymbol{u}_{1}^{t} + (\boldsymbol{u}_{2}^{t})^{T} \boldsymbol{K} (\phi) \boldsymbol{u}_{2}^{t},
\end{split}
\end{equation}
where $t$ is the designer selected time (e.g. $t = 2 \pi/5 s$, 9th time step) and $\boldsymbol{u}^{t}$ is the solution vector at the selected time. The optimized structures for this type of objective function are given in Figure \ref{fig:sin_cos_att_f1f2_1_0_0_final_design}. Unlike previous designs, the optimized designs evolve in a way that minimize the strain energy at the given time. The comparison of the strain energy profiles shows that the optimizer minimizes the strain energy at the given time, however, due to the nature of the defined objective function the strain energy after time $t$ is ignored by the optimizer. Of course this may be impractical, but here we simply wish to demonstrate the ability to influence the behavior of structure at a select time. This is shown in the strain energy profiles given in Figure \ref{fig:sin_cos_att_f1f2_1_0_0_final_design}. Moreover, the comparison of the performance of optimized designs under dynamic loading, via the strain energy profiles, shows the importance in the formulation of the objective function and timing, position, and magnitude of applied loads that need to be considered in the structural dynamics analysis, see Figure \ref{fig:compar_strain_energy_sm_ks_att}. 
%

%
\section{Conclusions
        \label{sec:Conclusions}}
The gradient-based topology optimization framework was presented to minimize the strain energy in  structures subjected to dynamic loading. The time-dependent sensitivities were provided via the adjoint method. The SIMP method was used to interpolate the stiffness of materials and HPM was used to improve the manufacturability and satisfy the minimum length scale in topology-optimized designs. Both static and dynamic optimization problems were considered in this paper. The optimized designs showed significant differences between structures designed for static and dynamic applied loads. Moreover, the optimized dynamic designs revealed the importance in the formulation of the objective function, timing and position of applied loads need to be considered in the structural dynamics design process. 
%

%
\section*{Acknowledgments
        \label{sec:Acknowledgments}}
This work was supported in part by the National Science Foundation under Grant No. 1538367.  This support is gratefully acknowledged.
%

%
\bibliographystyle{plain} 
\bibliography{share/JabRef_Database}

\end{document}